\documentclass[aps,twocolumn,superscriptaddress,amsmath]{revtex4}

\usepackage{epsf,latexsym}
\usepackage{amssymb,amsmath}
\usepackage{times,graphics}
\usepackage[final]{pdfpages}
\usepackage{framed}
\usepackage{epsfig,ulem}
\newcommand{\bra}[1]{\langle #1 |}
\newcommand{\ket}[1]{| #1 \rangle}

\def\6{\langle}
\def\9{\rangle}

\newcommand\cL{{\cal L}}
\newcommand\p{{\mathrm p}}
\newcommand\w{{\mathrm w}}
\newcommand\A{{\mathrm A}}
\newcommand\B{{\mathrm B}}
\newcommand\C{{\mathrm C}}
\newcommand\D{{\mathrm D}}

\newcommand\tr{{\mbox{Tr\,}}}

\def\half{\tfrac{1}{2}}

\newcommand{\ignore}[1]{}

\newcommand{\be}{\begin{equation}}
\newcommand{\ee}{\end{equation}}
\newcommand{\ba}{\begin{eqnarray}}
\newcommand{\ea}{\end{eqnarray}}

\begin{document}

\title{Is wave-particle objectivity compatible with determinism and locality?}
%\title{Determinism, locality and wave-particle objectivity are incompatible}
%\title{\green{Objectivity, determinism and locality: two may be compatible, three are not. }}
%\title{\green{Are locality and determinism compatible with wave-particle objectivity?}}
%\title{\red{Determinism,   in the delayed choice experiment}}
%Are determinism, locality and objectivity compatible?

\author{Radu Ionicioiu}
\affiliation{Department of Theoretical Physics, National Institute of Physics and Nuclear Engineering, 077125 Bucharest--M\u agurele, Romania}

\author{Thomas Jennewein}
\affiliation{Institute for Quantum Computing, University of Waterloo, Waterloo, Ontario, Canada}
\affiliation{Department of Physics and Astronomy, University of Waterloo, Waterloo, Ontario, Canada}

\author{Robert B. Mann}
\affiliation{Department of Physics and Astronomy, University of Waterloo, Waterloo, Ontario, Canada}
\affiliation{Perimeter Institute for Theoretical Physics, Waterloo, Ontario, Canada}

\author{Daniel R. Terno}
\affiliation{Department of Physics and Astronomy, Macquarie University, Sydney, NSW, Australia}

\begin{abstract}
Wave-particle duality, superposition and entanglement are among the most counterintuitive features of quantum theory. Their clash with our classical expectations motivated hidden-variable (HV) theories. With the emergence of quantum technologies we can test experimentally the predictions of quantum theory  {\em versus} HV theories and put strong restrictions on their key assumptions. Here we study an entanglement-assisted version of the quantum delayed-choice experiment and show that the extension of HV to the controlling devices only exacerbates  the contradiction. We compare HV theories that satisfy the conditions of objectivity (a property of photons being either particles or waves, but not both), determinism, and local independence of hidden variables with quantum mechanics. Any two of the above conditions are compatible with it. The conflict becomes manifest when all three conditions are imposed and persists for any non-zero value of entanglement. We propose an experiment to test our conclusions.

\end{abstract}

\maketitle

%\section*{Introduction}

\noindent Quantum mechanics is proverbially counterintuitive \cite{qf1,peres}. For many years thought experiments were used to dissect its puzzling properties, while hidden variable (HV) models strived to explain, or even to remove them \cite{qf1,peres,branyan,eng13}. The development of quantum technologies \cite{qinfo1,qinfo2} enabled us not only to perform several former {\it gedanken} experiments \cite{qf1,peres}, but also to devise new ones \cite{sew,qbohrein,qcon1,qcon2,it11}. One can gain new insights into quantum foundations by introducing quantum controlling devices \cite{qcon2,it11,six14} into well-known experiments. This has led, for example, to a reinterpretation \cite{it11,qcomp,qur-comp} of Bohr's complementarity principle \cite{bohr}.

\begin{figure}[h!]
\includegraphics[width=0.48\textwidth]{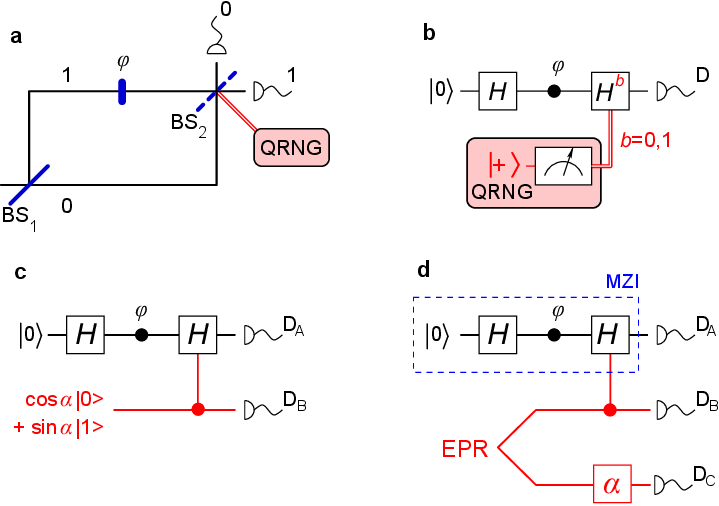}
\caption{\sf \textbf{The evolution of the delayed-choice experiment.} (\textbf{a}) In Wheeler's classic experiment, the second beam-splitter is inserted or removed after the photon is inside the interferometer; this prevents the photon from changing its mind \cite{wheeler_wz} about being a particle or a wave. The detectors observe either an interference pattern depending on the phase $\varphi$ (wave behaviour), or an equal distribution of hits (particle behaviour). A quantum random number generator (QRNG) determines whether BS$_2$ is inserted or not.\\
Quantum networks: (\textbf{b}) in the classic delayed-choice experiment the QRNG is an auxiliary quantum system initially prepared in the equal superposition state $\ket{+}=\tfrac{1}{\sqrt{2}}(\ket{0}+ \ket{1})$ and then measured. The Hadamard gate $H$ is the quantum network equivalent of the beamsplitter; (\textbf{c}) delayed-choice with a quantum control \cite{it11}; (\textbf{d}) entanglement-assisted quantum delayed-choice experiment \cite{kaiser12}. The ancilla $\C$ is measured along the direction $-\alpha$, equivalent to the application of a rotation $R_y(\alpha)= e^{i\alpha \sigma_y}$ before a measurement in the computational basis.}
\label{struc}
\vspace{-5mm}
\end{figure}

Wave-particle duality is best illustrated by the classic Wheeler delayed-choice experiment (WDC) \cite{wheeler_wz, leggett, exp_dc}, Figure \ref{struc}(a-b). A photon enters a Mach-Zehnder interferometer (MZI) and its trajectory is coherently split by the beamsplitter BS$_1$ into an upper and a lower path. The upper path contains a variable phase shift $\varphi$. A random number generator controls  {the insertion ($b=1$) or
removal ($b=0$) of a second beamsplitter BS$_2$}. If BS$_2$ is present, the interferometer is closed and we observe an interference pattern depending on the phase shift $\varphi$. If BS$_2$ is absent, the MZI is open and the detectors measure a constant probability distribution independent of $\varphi$. Thus, depending on the experimental setup, the photon behaves in two completely different ways. In the case of the closed MZI, the interference pattern suggests that the photon traveled along both paths simultaneously and interfered with itself at the second beamsplitter BS$_2$, hence showing a wavelike behaviour. However, if the interferometer is open, since always only one of the two detectors fires, one is led to the conclusion that the photon traveled only one path, hence displaying a particle-like behaviour.

The complementarity of  the interferometer setups required to observe {particle or  wave behaviour} obscures the simultaneous presence of both properties, allowing the (objective) view that, at any moment of time, a photon  can be either a particle or a wave. The WDC experiment uncovers the difficulty inherent in this view by randomly choosing whether or not to insert the second beamsplitter (BS$_2$) after the photon enters the interferometer (Figure \ref{struc}a). This delayed choice prevents a possible causal link between the experimental setup and the photon's behaviour: the photon should not know beforehand if it has to behave like a particle or like a wave.

The delayed-choice experiment with a quantum control (Figure \ref{struc}c) highlights the complexity of space-time ordering of events, once parts of the experimental setup become quantum systems \cite{it11}. The quantum-controlled delayed-choice experiment has been recently implemented in several different systems \cite{qdc_NMR1, qdc_NMR2, qdc_optics, job, kaiser12}. In order to ensure the quantum behaviour of the controlling device one can either test the Bell inequality \cite{job} or use an entangled ancilla \cite{kaiser12}.

The theoretical analysis of the quantum WDC involved so far a single binary hidden-variable $\lambda$ describing the classical concepts of wave/particle. Here we introduce a full HV description for both the photon $\A$ and the ancilla. We analyze the relationships between the concepts of determinism,  wave-particle objectivity and local independence of HV in the entanglement-controlled delayed-choice experiment. We show that, when combined, these assumptions lead to predictions that are different from those of quantum mechanics, even if any two of them are compatible with it. We propose and discuss an experiment to test our conclusions.

\medskip

~\\
{\sf \textbf{Results}}
%\section*{Results}

\noindent\textbf{Notation.}
We use the conventions of \cite{six14, branyan}; $q(a,b,\ldots)$ are the quantum-mechanical probability distributions and $p(a,b, \ldots,\varLambda)$ the predictions of HV theories with a hidden variable $\varLambda$. We consider either a single HV $\varLambda$ which  fully determines  behaviour of the system, or refine it as $\varLambda_1$, $\varLambda_2$ pertaining to different parts of the system. For simplicity we assume $\varLambda$ is discrete; the analysis can be easily generalized to the continuous case.

\noindent\textbf{Quantum system.}
The system we analyze consists of three qubits: a photon A and an entangled pair BC (Fig.~\ref{struc}d). We denote the measurement outcomes for the photon A as $a=0,1$, and for the two ancilla qubits as $b$ and $c$; the corresponding detectors are $\D_\A$, $\D_\B$ and $\D_\C$. The system is prepared in the initial state $\ket{0}_\A \big( \sqrt{\eta}\ket{00}+ \sqrt{1-\eta}\ket{11} \big)_{\B\C}$; for $\eta=\half$, BC is a maximally entangled EPR pair.

Photon A enters a Mach-Zehnder interferometer in which the second beamsplitter is quantum-controlled by qubit B. The third qubit C undergoes a $\sigma_y$ rotation $R_y(\alpha)= e^{i\alpha \sigma_y}$ followed by a measurement in the computational basis. The state before the measurements is
\ba
\ket{\psi}&= \big( \sqrt{\eta} \cos\alpha \ket{{\p}} \ket{0} + \sqrt{1-\eta} \sin\alpha \ket{{\w}} \ket{1} \big)_{\A\B} \ket{0}_\C \nonumber \\
&-  \big( \sqrt{\eta} \sin\alpha \ket{{\p}} \ket{0} - \sqrt{1-\eta} \cos\alpha \ket{{\w}} \ket{1} \big)_{\A\B} \ket{1}_\C.
\label{output}
\ea
The counting statistics that result from the particle-like state  $\ket{{\p}}= \tfrac{1}{\sqrt{2}} (\ket{0}+ e^{i\varphi}\ket{1})$ and the wave-like state $\ket{{\w }}= e^{i\varphi/2}(\cos\tfrac{\varphi}{2}\ket{0}- i \sin\tfrac{\varphi}{2}\ket{1})$ are discussed below (Eqs.~\eqref{p_and_w.1}, \eqref{p_and_w.2} and Methods).

%hv info
\noindent{\bf Constraints on HV theories.} Our strategy is to show that $q(a, b, c)$ cannot result from a probability distribution $p(a, b, c, \varLambda)$ of a hidden-variable theory satisfying the requirements of  wave-particle objectivity, local independence and determinism. Any viable HV theory should satisfy the adequacy condition: namely it should reproduce the quantum statistics by summing over all hidden variables $\varLambda$:
\be
q(a, b, c)= p(a, b, c):= \sum_{\varLambda} p(a, b, c, \varLambda).
\label{qp}
\ee

We encapsulate the additional classical expectations into three assumptions (see Box 1 for the formal definitions of the concepts we consider in this section).

\begin{framed}
\sf \small
\noindent{\textbf{Box 1 $\vert$ Three classical assumptions}}

\noindent{ {\textbf{Wave-particle objectivity.}}} We define particles and waves according to the experimental behaviour in an open, respectively closed, Mach-Zender interferometer \cite{it11}. A particle in an open interferometer ($b=0$) is insensitive to the phase shift in one of the arms and therefore has the statistics
\be
p(a|b=0, \varLambda)= \big(\half, \half \big), \ \ \ \forall \varLambda \in \cL_\p.
\label{p_and_w.1}
\ee
By contrast, a wave in a closed MZI ($b=1$) shows interference
\be
p(a| b=1, \varLambda)= \big(\cos^2\!\tfrac{\varphi}{2}, \sin^2\!\tfrac{\varphi}{2}\big), \ \ \ \forall \varLambda \in \cL_\w.
\label{p_and_w.2}
\ee
The sets $\cL_\p$ and $\cL_\w$ must be disjoint; otherwise there are values of $\varLambda$ that introduce wave-particle duality. Writing  $\cL_\p \cup \cL_\w=\cL$, the wave/particle property is expressed by a mapping $\lambda: \cL \mapsto \{ \p, \w \}$ and the sets $\cL_\p = \lambda^{-1}(\p)$, $\cL_\w = \lambda^{-1}(\w)$ are the pre-images of $\p, \w$ under the function $\lambda$.

\noindent \textbf{Determinism.} The hidden variable $\varLambda$ determines the individual outcomes  of the detection \cite{branyan}. Specifically, for the setup of (Fig.~1d)
\be
p(a,b,c|\varLambda)= \chi_{abc}(\varLambda), \label{cf}
\ee
where the indicator function $\chi=1$, if $\varLambda$ belongs to some pre-determined set, and $\chi=0$ otherwise.

\noindent  \textbf{Local independence.} The HV $\varLambda$ are split into $\varLambda_1$ and $\varLambda_2$, and the prior probability distribution has a product structure
\be
p(\varLambda)= f(\varLambda_1) F(\varLambda_2),  \label{HV-local}
\ee
for some probability distributions $f$ and $F$, where the subscripts 1 and 2 respectively refer to the photon A and the pair BC. Such bilocal variables have been previously considered \cite{bilocal}.
\end{framed}

For a given photon we require the property of being a particle or a wave to be  objective (intrinsic), that is, to be unchanged during its lifetime. This condition selects from the set of adequate HV theories those models that have meaningful notions of particle and wave \cite{it11}. For each photon, the hidden variable $\varLambda$ should determine unambiguously if the photon is a particle or a wave,  thus allowing the partition of the set of hidden variables $\cL$ into two disjoint subsets, $\cL = \cL_\p \cup \cL_\w$, where the subscript indicates the property, particle or wave.

The particle (wave) properties are abstractions of the particle (wave) counting statistics in open (closed) MZI, respectively.
%Equations %\eqref{p_and_w.1},\eqref{p_and_w.2} hold for individual $\varLambda$'s (and not statistically), since being a particle (wave) holds individually for %a given photon.
The behaviour of a particle (wave) in a closed (open) MZI is not constrained; this allows for significant freedom in constructing HV theories.  Experimentally, the wave or particle behaviour  depends only on the photon and the settings of the MZI:
\be
{p(a|b,c,\varLambda)= p(a|b,\varLambda)},
\label{cindep}
\ee
for all values of $a$, $b$, $c$ and $\varLambda$.

By replacing the single qubit ancilla with an entangled pair one can take advantage of both the quantum control and the space-like separation between events.  The rationale behind the third qubit C is that it allows us to choose the rotation angle $\alpha$ after both qubits A (the photon) and B (the quantum control) are detected. This is not possible in the standard quantum WDC \cite{it11}, Fig.\ref{struc}c, where the quantum control B has to be prepared (by setting the angle $\alpha$) before it interacts with A. As discussed in Methods, there is a unique assignment of probabilities that satisfies all the requirements of adequacy, wave-particle objectivity and determinism. Adopting this assignment we reach the same level of incongruity as in \cite{it11}, since the probability $p(\lambda)$ of  photon A being a particle or a wave is determined by the entanglement between B and C,
\be
p(\lambda)= (\eta,1-\eta).
\label{lastcon}
\ee

This incongruity becomes an impossibility when the photon A and the  entangled pair BC are prepared independently. In this case their hidden variables are generated independently; that is, a single HV $\varLambda$ not only has the structure $\varLambda= (\varLambda_1,\varLambda_2)$, where the subscripts 1 and 2 refer to the photon A and the pair BC, respectively, but the prior probability distribution of HV has a product form. To realize this condition experimentally we rely on the absence of the superluminal communication and a space-like separation of the two events.

Unlike the typical Bell-inequality scenarios we have a single measurement setup which involves two independent HV distributions. Moreover, by performing the rotation $R_y(\alpha)$ and the detection $\D_\C$ sufficiently fast, such that the information about A and $\varLambda_1$ cannot reach the detector $\D_\C$, the detection outcome is determined only by $\varLambda_2$. Since being a wave (particle) is assumed to be an objective property of A, $\lambda= \lambda(\varLambda_1)$ is a binary function of the HV $\varLambda_1$ only.

\noindent{\bf Contradiction.} We show in Methods that for $\eta \neq 0,1$ (these two cases correspond to an always closed or open MZI) the requirements of adequacy, wave-particle objectivity, determinism and local independence are satisfied only if
\be
 \cos 2\alpha=0.
\label{contra1}
\ee
This proves our main theoretical result: determinism, local independence and  wave-particle objectivity are not compatible with quantum mechanics for any $\alpha\ne {\pm\pi/4, \pm3\pi/4}$. We will later discuss  how exactly a HV theory that satisfies the three classical assumptions is inadequate.

%experiment

\begin{figure}[htbp]
\includegraphics[width=0.48\textwidth]{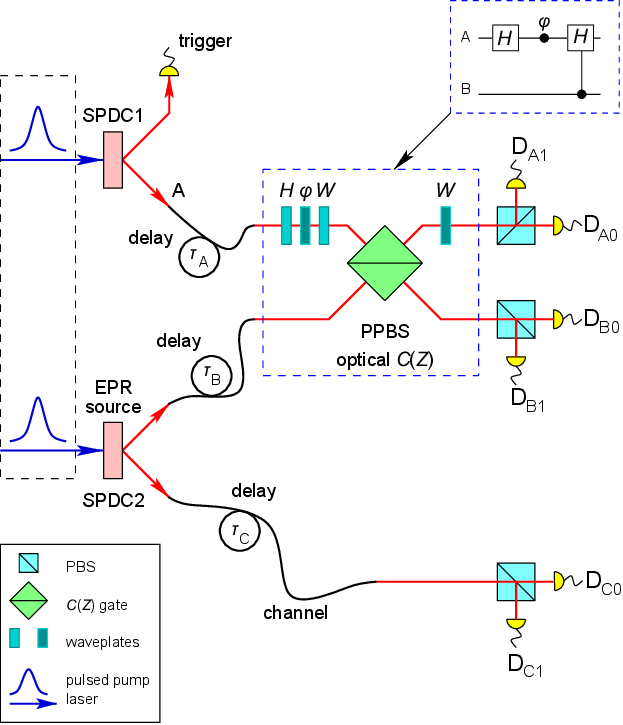}

\caption{\sf \textbf{Proposed experimental setup.} Two spacelike separated pump pulses (blue) generate, via SPDC, two pairs of entangled photons (red). The first photon is the trigger and the  other three the photons $\A,\B,\C$. Inset: the quantum-controlled Mach-Zehnder interferometer. The optical delays in the three photon arms, $\tau_\A, \tau_\B, \tau_\C$ can be adjusted to ensure the desired time ordering of the detection events.}
\label{setup}
\end{figure}

\noindent{\bf Proposed experiment.}  In figure \ref{setup} we show the proposed experimental setup for the entanglement-controlled delayed-choice experiment. Two pump pulses (blue) are incident on two nonlinear crystals and generate via spontaneous parametric down-conversion (SPDC) two pairs of entangled photons (red). One of the photons is the trigger and the other three are the photons A, B, C, with BC being the entangled pair.

Photons A and B are held in the lab (with appropriate delay lines) and together they implement the controlled MZI. The central element is the quantum switch, which is the controlled-Hadamard gate $C(H)= (W\otimes I) C(Z) (W\otimes I)$, where $W= \sigma_z e^{i\tfrac{\pi}{8} \sigma_y}$. The photonic controlled-$Z$ gate $C(Z)$ is implemented with a partially-polarizing beam-splitter (PPBS) and is done probabilistically via post-selection \cite{ppbs_cnot1,ppbs_cnot2}. Optical wave-plates perform single-qubit rotations (gates $H$, $\varphi$ and $W$) on photon A. Photon C is sent through a channel at a distant location, then measured in a rotated basis. Two independent lasers generate the two photon pairs (Fig.\ref{setup} \cite{indep_lasers, ent_swap}); in this case we can use Eq.~\eqref{HV-local} to describe independent probability distributions for $\varLambda_1$ and $\varLambda_2$.

\begin{figure}
    \includegraphics[width=\columnwidth]{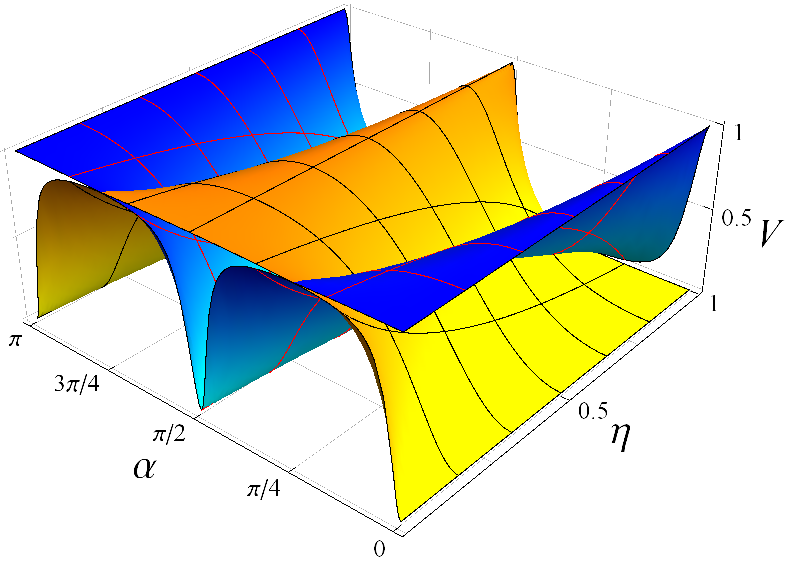}
    \setlength{\unitlength}{0.01\linewidth}
    \begin{picture}(100,0)    % picture environment for inset
        \put(68,64){\includegraphics[width=0.125\textwidth]{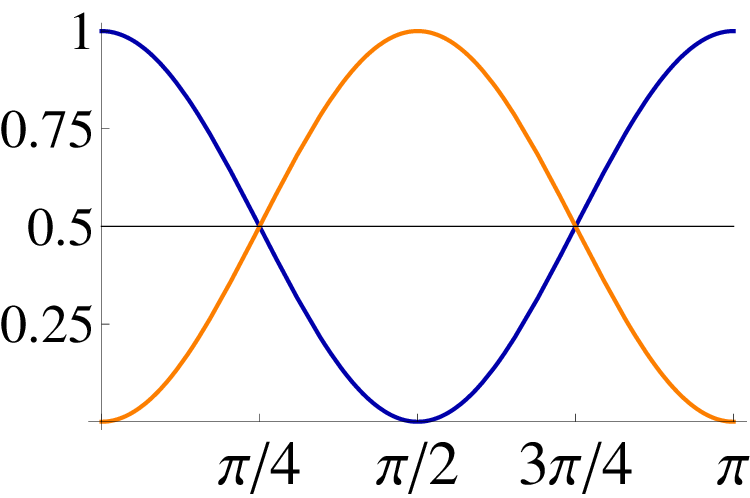}}
    \end{picture}
    \caption{\sf \textbf{Visibility.} The visibilities $V_{\A|c=0}$ (yellow) and $V_{\A|c=1}$ (blue), are calculated in Methods. In HV theories the visibility does not distinguish between the $c=0,1$ cases. The inset illustrates this for $p(\p)=\eta=\frac{1}{2}$, with the straight line representing the HV visibility prediction.}
\label{visib}
\vspace{-3mm}
\end{figure}

~\\
{\sf \textbf{Discussion}}
%\section*{Discussion}

\noindent
In this section we consider how exactly a HV theory, which satisfies the three classical assumptions, fails the adequacy test. The interference pattern measured by the detector $\D_{\A0}$ is $I_\A(\varphi)= \tr(\rho_\A \ket{0}\bra{0})$, with $\rho_\A= \tr_{\B\C}\ket{\psi}\bra{\psi}$ the reduced density matrix of photon A. The data can be postslected according to the outcome $c$ resulting in $I_{\A|c}$. The visibility of the interference pattern (Methods) is $V= (I_{\max}- I_{\min})/(I_{\max}+ I_{\min})$, where the min/max values are calculated with respect to $\varphi$. The postselected visibility for $c=0$ is (Fig.\ref{visib})
\be
V_{\A|c=0}= \frac{(1-\eta) \sin^2\!\alpha}{\eta\cos^2\!\alpha+(1-\eta) \sin^2\!\alpha}
\ee
The full (non-postselected) visibility is $V_A= 1- \eta$ and gives information about the initial entanglement of the BC pair. On the other hand, if one assumes that the HV are distributed according to Eq.~\eqref{HV-local} and satisfy the wave-particle objectivity and determinism, the visibility is independent of $c$,
\be
V_\A^{\mathrm{HV}}\equiv V_{\A|c=0}^{\mathrm{HV}}=V_{\A|c=1}^{\mathrm{HV}}= 1-f,
\ee
in contrast with the quantum-mechanical prediction (Figure \ref{visib}). Details of this calculation are in the Methods.

This incompatibility between the basic tenets of hidden-variable theories  and quantum mechanics has two remarkable features. First, the contradiction is revealed for  any, arbitrarily small, amount of entanglement. This test is in sharp distinction with Bell-type experiments insofar as our result is free from inequalities. Wave-particle objectivity, revealed only statistically, is more intuitive and technically milder than the assumption of sharp values of quantum incompatible observables. Second, in our set-up any two of the classical ideas together are compatible with the quantum-mechanical predictions. This  fact, and the way we arrived  at the contradiction, invite questions concerning the internal consistency of classical concepts \cite{imt14}.

~\\
{\sf \textbf{Methods}}
%\section*{\bf Methods}

\small
\noindent{\sf \textbf{\footnotesize{Quantum-mechanical analysis.}}}
The initial state of photons A, B and C is
\be
\ket{0}_\A \big( \sqrt{\eta}\ket{00}+ \sqrt{1-\eta}\ket{11} \big)_{\B\C}.
\ee
The ancilla qubits B and C are maximally entangled for $\eta= \half$. The final state before measurement is given by Eq.~\eqref{output}. From it we calculate the quantum statistics $q(a, b, c)$, where each of $a$, $b$, and $c$ take the values $\{0,1\}$. The probability distribution for $c=0$ is
\begin{align}
q(a, b, c=0)= \big(& \half \eta \cos^2\alpha, (1-\eta) \sin^2\alpha \cos^2 \tfrac{\varphi}{2}, \nonumber \\
& \half \eta \cos^2\alpha, (1-\eta) \sin^2\alpha \sin^2 \tfrac{\varphi}{2} \big).   \label{c0q}
\end{align}
where the four entries correspond to the values $(a,b)= (00, 01, 10, 11)$. For $c=1$ we obtain
\begin{align}
q(a, b, c=1)=\big(& \half \eta \sin^2\alpha, (1-\eta) \cos^2\!\alpha \cos^2 \tfrac{\varphi}{2}, \nonumber \\
 &\half \eta \sin^2\!\alpha, (1-\eta) \cos^2\!\alpha \sin^2 \tfrac{\varphi}{2} \big). \label{c1q}
 \end{align}
This in turn yields
\begin{align}
& q(a,b) = ( \half \eta, (1-\eta) \cos^2 \tfrac{\varphi}{2}, \half \eta, (1-\eta) \sin^2 \tfrac{\varphi}{2} \big), \\
& q(b,c) = (\eta\cos^2\! \alpha, \eta\sin^2 \!\alpha, (1-\eta)\sin^2 \!\alpha, (1-\eta)\cos^2\! \alpha ),\\
& q(b) = (\eta, 1-\eta),  \label{qb} \\
& q(c) = (\eta\cos^2\!\alpha+ (1-\eta) \sin^2\!\alpha, \eta\sin^2\!\alpha+ (1-\eta) \cos^2\!\alpha). \label{qc}
\end{align}
For $\eta = \half$ the probability distributions for $b$ and $c$ are equal. If $\eta \ne \half$, B and C are no longer maximally entangled and the symmetry between them is broken: a rotation $\alpha$ on C no longer corresponds to a rotation $\alpha$ on B.
The conditional probabilities are
\ba
q(c|b)= (\cos^2 \alpha, \sin^2 \alpha, \sin^2 \alpha, \cos^2 \alpha),
\ea
and from Bayes' rule $q(b|c)= q(c|b) q(b)/q(c)$.

\medskip

\noindent{\sf \textbf{\footnotesize{Solution to the three constraints.}}}
We now show that it is possible to construct a HV model that is adequate, objective and deterministic. The unknown parameters at our disposal are 16  probabilities $p(a,b,c,\lambda)$. These probabilities are derived from the underlying distribution $p(\varLambda)$ summed over appropriate domains. At this stage we do not enquire about the connection with the HV $\varLambda$. The probabilities $p(a,b,c,\lambda)$ satisfy seven adequacy constraints, Eqs.~\eqref{c0q} and \eqref{c1q}, plus the normalization constraint. The adequacy conditions can be written as
\be
q(a,b,c)=p(a,b,c)=p(a,b,c,\p)+p(a,b,c,\w).
\label{adeabc}
\ee

In addition,  Eq.~\eqref{cindep} and the standard rules for the conditional probabilities, such as
\be
p(a|b,\lambda)\equiv p(a|b,c,\lambda)=\frac{p(a,b,c,\lambda)}{p(0,b,c, \lambda)+ p(1,b,c, \lambda)},
\ee
imply the existence of four additional constrains,
\begin{align}
& p(0,0,c,\p)=p(1,0,c,\p), \\
& p(0,1,c,\w)\sin^2\tfrac{\varphi}{2}= p(1,1,c,\w)\cos^2\tfrac{\varphi}{2}. \label{drealism-p}
\end{align}

The resulting linear system has a four-parametric family of solutions. However, a straightforward calculation shows that  for all these solutions $p_4(a,b,c,\lambda)$ the resulting statistics in an open/closed MZI is independent of $\lambda$,
\begin{align}
& p_4(a|b=0,\p)=p_4(a|b=0,\w) =\big(\half,\half), \\
& p_4(a|b=1,\w)=p_4(a|b=1,\p) =\big(\cos^2\!\tfrac{\varphi}{2}, \sin^2\!\tfrac{\varphi}{2}\big),
\end{align}
that is, the statistics of $\D_\A$ is determined solely by the state of the interferometer.

We can avoid the reintroduction of wave-particle duality using a special solution
\be
p_{\mathrm s}(b| \lambda)= \delta_{\lambda \p}\delta_{b0} +\delta_{\lambda \w}\delta_{b1}\equiv p_{\mathrm s}(\lambda| b),
\label{blambda}
\ee
which imposes  the $b$-$\lambda$ correlation (compare \cite{it11}). As a result,
\be
p(b=0,\lambda=\w)=p(b=1,\lambda=\p)=0,
\ee
and since the probabilities are positive,
\be
\sum_{a,c}p(a,1,c,\p)=\sum_{a,c}p(a,0,c,\w)=0,
\ee
the eight above probabilities are zero individually. The system appears overconstrained, but it still has a unique solution
\be
p_{\mathrm s}(a,b,c,\lambda)=q(a,b,c)p_{\mathrm s}(b|\lambda).
\ee
In particular,
 \be
p_{\mathrm s}(\lambda)= \sum_{a,b,c} p_{\mathrm s}(a,b,c,\lambda)= (\eta, 1-\eta). \label{etalambda}
\ee

\medskip

\noindent{\sf \textbf{\footnotesize{Deriving the contradiction.}}}
In addition to the partition of $\cL$ according to the values of $\lambda=\p,\w$ we will use the decomposition of the set of HV according to the outcomes of $\D_\C$. The two branches $c=0,1$ correspond to the partition
\be
\cL=\cL_0\cup\cL_1,
\ee
where for $\varLambda\in\cL_c$ the outcome of $\D_\C$ is $c$. The assumption of local independence implies a Cartesian product structure
\be
\cL=\{\varLambda_1\}\times \big(\cL_0^2\cup\cL_1^2\big)=\big(\cL_\p^1\cup\cL^1_\w\big)\times \big(\cL_0^2\cup\cL_1^2\big),
\ee
of the set of HV, where the subsets depend on the experimental set-up. When the superscripts 1 and 2 on $\cL$ are redundant, we may omit them.

Now we show that under the assumptions of adequacy and the three classical assumptions of  wave-particle objectivity, determinism and local independence it is impossible to derive the  solution $p(a,b,c,\lambda)$ with any arrangement of  the probabilities $p(\varLambda)$. The probability of the outcome $c$ satisfies
\be
q(c)\equiv p(c)= \sum_{\varLambda\in\cL_c} p(\varLambda)= \sum_{\varLambda\in\cL_\p\cap\cL_c} p(\varLambda)+\sum_{\varLambda\in\cL_\w\cap\cL_c} p(\varLambda).
\label{c-prob}
\ee
To simplify the calculations we enumerate the variables $\varLambda_{1,2}$ by the indices $i,j$, respectively. The domain $\cL_c^2$ corresponds, according to the hypothesis, to the index set $J_c$ of $\varLambda_2$, and the domains $\cL^1_\p$ and $\cL^1_\w$ to the index sets $I_\p$ and $I_\w$ of $\varLambda_1$, respectively. In particular,
\be
p(\lambda)=\left(\sum_{i\in I_\p}f_i,\sum_{i\in I_\w}f_i\right), \label{lambda-loc}
\ee
for some $f_i\equiv f(\varLambda_1^i)$. The prior distribution  of HV and the domains of summation can depend on the parameters $\eta$, $\varphi$, and $\alpha$.

The putative behaviour of a wave ($\lambda=\w$) in an open ($b=0$) interferometer and of a particle ($\lambda=\p$) in a closed ($b=1$) one is characterized by two unknown distributions  $x_{ij},i\in I_\w$ and $y_{i  j},i\in I_\p$, respectively
\begin{align}
&p\big(a|b=0, \varLambda= (\varLambda_1^i, \varLambda_2^j)\big) = (x_{i  j}, 1-x_{i  j}), \qquad i\in I_\p  \\
&p\big(a|b=1, \varLambda= (\varLambda_1^i, \varLambda_2^j)\big) =(y_{i  j}, 1-y_{i  j}), \qquad i\in I_\w
\end{align}
allowing for a possible dependence on a value of $\varLambda_2$. The remaining two sets of variables are the probability distributions for $b$ conditioned on the values of hidden variables $\varLambda$:
\begin{align}
&p\big(b| \varLambda= (\varLambda_1^i, \varLambda_2^j)\big) = (z_{i  j}, 1-z_{i  j}), \qquad i\in I_\p,   \label{b-data1}\\
&p\big(b| \varLambda= (\varLambda_1^i, \varLambda_2^j)\big) =(v_{i  j}, 1-v_{i  j}), \qquad i\in I_\w.  \label{b-data2}
\end{align}

The requirement of adequacy means that the proposed HV theory reproduces the quantum statistics given above. For compactness we refer to the probability of having the HV values $(\varLambda_1=\varLambda_1^i, \varLambda_2=\varLambda_2^j)$, $p(\varLambda_1^i,\varLambda_2^j)$, as $p_{ij}$, using the same convention as for $x_{ij}$, $y_{ij}$, $z_{ij}$ and $v_{ij}$. For $c=0$ we have
\begin{widetext}
\ba
&& q(0,0,0)= \half \eta\cos^2\!\alpha\equiv\half \sum_{i\in I_\p, j\in J_0} z_{ij} p_{ij}+ \sum_{i\in I_\w, j\in J_0}x_{ij}v_{ij} p_{ij}, \label{00} \\
&& q(0,1,0)= (1-\eta) \sin^2\!\alpha \cos^2 \tfrac{\varphi}{2}\equiv \sum_{i\in I_\p, j\in J_0}y_{ij} (1-z_{ij}) p_{ij}+ \cos^2 \tfrac{\varphi}{2}\sum_{i\in I_\w, j\in J_0}(1-v_{ij}) p_{ij} \label{01},\\
&& q(1,0,0)= \half \eta\cos^2\!\alpha\equiv\half \sum_{i\in I_\p, j\in J_0} z_{ij}p_{ij}+
\sum_{i\in I_\w, j\in J_0} (1-x_{ij})v_{ij}p_{ij}, \label{10} \\
&& q(1,1,0)= (1-\eta) \sin^2\!\alpha \sin^2 \tfrac{\varphi}{2}\equiv  \sum_{i\in I_\p, j\in J_0}(1-y_{ij}) (1-z_{ij}) p_{ij}+ \sin^2 \tfrac{\varphi}{2}\sum_{i\in I_\w, j\in J_0}(1-v_{ij}) p_{ij}, \label{11}
\ea
\end{widetext}
with analogous expressions for $c=1$.  Adding and subtracting Eqs.~\eqref{00} and \eqref{10} we obtain, respectively
\ba
&& \sum_{i\in I_\p, j\in J_0} z_{ij} p_{ij}+\sum_{i\in I_\w, j\in J_0}v_{ij} p_{ij}=\eta\cos^2\alpha \\
&& \sum_{i\in I_\w, j\in J_0} (1-2x_{ij})v_{ij}p_{ij}=0
\ea

Adding Eqs.~\eqref{01} and \eqref{11} yields
\be
(1-\eta) \sin^2\!\alpha= \sum_{i\in I_\p, j\in J_0}  (1-z_{ij}) p_{ij}+\sum_{i\in I_\w, j\in J_0}(1-v_{ij}) p_{ij}    ,
\ee
which upon substitution back into Eq.~\eqref{01} results in
\be
\sum_{i\in I_\p, j\in J_0}\left(\cos^2 \tfrac{\varphi}{2}-y_{ij}\right)(1-z_{ij})p_{ij}=0.
\ee

Four additional equations (giving a total of seven independent equations) are obtained for $j\in J_1$ with $\cos^2\alpha\leftrightarrow \sin^2\alpha$.

\medskip

From Eq.~\eqref{blambda} it follows that $v_{ij}=0, i\in I_\w$ and $z_{ij}=1, i\in I_\p$. Hence for $c=0$ only two equations are not automatically satisfied,
\be
\eta\cos^2\!\alpha=\sum_{i\in I_\p, j\in J_0}p_{ij}, \qquad (1- \eta)\sin^2\!\alpha=\sum_{i\in I_\w, j\in J_0}p_{ij} . \label{non-1}
\ee
The corresponding equations for $c=1$ are
\be
\eta\sin^2\!\alpha=\sum_{i\in I_\p, j\in J_1}p_{ij}, \qquad (1- \eta)\cos^2\!\alpha=\sum_{i\in I_\w, j\in J_1}p_{ij}, \label{non-2}
\ee
which are in agreement with $q(c)$, Eq.~\eqref{qc}.

Now we use the product structure of the probability distribution, Eq.~\eqref{HV-local},
\be
p(\varLambda)= f(\varLambda_1)F(\varLambda_2) \Leftrightarrow p_{ij}=f_i F_j.
\ee
Using Eq.~(28) and Eq.~\eqref{etalambda} we find that
\be
\sum_{i\in I_\p}f_i=\eta. \label{eta-f}
\ee
Adding the pairs of equations in \eqref{non-1} and \eqref{non-2} and summing over the index $i$ we express the adequacy condition
 $q(c)=\sum_{j\in I_c} F_j$,
\begin{align}
\eta\cos^2\!\alpha+ (1-\eta) \sin^2\!\alpha= \sum_{j\in J_0}F_j, \\
 \eta\sin^2\!\alpha+ (1-\eta) \cos^2\!\alpha= \sum_{j\in J_1}F_j,
\end{align}
but on the other hand, for $\eta\neq0,1$ summing over $i$ in each of these four equations separately and using Eq.~\eqref{eta-f} we get
\be
\sum_{j\in J_0}F_j=\cos^2\alpha=\sin^2\alpha=\sum_{j\in J_1}F_j.
\ee
These equations can be satisfied for any $\eta$ only if
\be
\cos^2\alpha=\sin^2\alpha,
\ee
resulting in the contradiction (for arbitrary $\alpha$) $\cos 2\alpha =0$.

\medskip

\noindent{\sf \textbf{\footnotesize{Experimental signature.}}} The interference pattern measured by the detector $\D_{\A}$ is $I_\A(\varphi)= \tr(\rho_A \ket{0}\bra{0})$, with $\rho_A= \tr_{BC}\ket{\psi}\bra{\psi}$ the reduced density matrix of photon A. The data can be postselected according to the outcome $c$ resulting in $I_{\A|c}$. The intensity (signal) measured by detector $\D_{\A}$ for $c=0$ (and no post-selection on $b$) is:
\be
I_{\A|c=0}= \half \eta\cos^2\!\alpha+(1-\eta) \sin^2\!\alpha \cos^2 \tfrac{\varphi}{2},
\ee
giving the visibility
\be
V_{\A|c=0}= \frac{(1-\eta) \sin^2\!\alpha}{\eta\cos^2\!\alpha+(1-\eta) \sin^2\!\alpha}.
\ee
A similar calculation gives the visibility for $c=1$
\be
V_{\A|c=1}= \frac{(1-\eta) \cos^2\!\alpha}{\eta\sin^2\!\alpha+(1-\eta) \cos^2\!\alpha}.
\ee

The full intensity measured by detector $\D_{\A}$ (without postselecting on $c$) is $I_\A= \half \eta+ (1-\eta) \cos^2 \tfrac{\varphi}{2}$ and the corresponding visibility
\be
V_\A= 1- \eta.
\ee
Thus the visibility of detector $\D_{\A}$ gives information about the entanglement of the BC pair.

We now calculate the visibilities predicted by a non-trivial HV theory that is assumed to satisfy the three classical assumptions. Using Eq.~\eqref{blambda} we rewrite the counting statistics as
\begin{align}
p(0,0,0) & = \half \sum_{i\in I_\p, j\in J_0} f_i F_j, \\
p(0,0,1) &= \half \sum_{i\in I_\p, j\in J_1} f_i F_j     \label{p01}  \\
p(0,1,0) & =   \cos^2\! \tfrac{\varphi}{2} \sum_{i\in I_\w, j\in J_0} f_i F_j, \\
p(0,1,1) & = \cos^2\! \tfrac{\varphi}{2}\sum_{i\in I_\w, j\in J_0} f_i F_j.    \label{p11}
\end{align}
For the product probability distribution above we get
\begin{align}
p(0,0|j) &= \frac{p(0,0,j)}{p(c=j)} = \frac{\half \sum_{i\in I_\p, k\in J_j} f_i F_k}{\sum_{k\in J_j}  F_k} = \half f \\
p(0,1|j) &= \frac{p(0,1,j)}{p(c=j)} = \frac{\cos^2\! \tfrac{\varphi}{2}\sum_{i\in I_\w, k\in J_j} f_i F_k}{\sum_{k\in J_j}  F_k} = \cos^2\! \tfrac{\varphi}{2}\,(1-f)
\end{align}
for $j=0,1$ separately, where $f=\sum_{i\in I_p}f_i$.  As a result,
\be
I^{\mathrm{HV}}_{\A|c=0} {=I^{\mathrm{HV}}_{\A|c=1}}= \half f+ \cos^2\! \tfrac{\varphi}{2}\,(1-f)
\ee
giving
\be
 V^{\mathrm{HV}}_{\A|c=0} {= V^{\mathrm{HV}}_{\A|c=1}}= 1- f,
\ee
for the visibilities in HV theories.

\medskip

\noindent {\bf Acknowledgments}\\
\small
D.R.T. thanks Perimeter Institute for support and hospitality. We thank Lucas C\'eleri, Jim Cresser, Berge Englert, Peter Knight, Stojan Rebi\'c, Valerio Scarani, Vlatko Vedral and Man-Hong Yung for discussions and critical comments, and Alla Terno for help with visualization. This work was supported in part by the Natural Sciences and Engineering Research Council of Canada. R.I.~acknowledges support from the Institute for Quantum Computing, University of Waterloo, Canada, where this work started.

\medskip

%\normalsize
\noindent{\bf Author contributions}\\
\small
\noindent T.J. and R.I. conceived the entanglement-controlled protocol.  D.R.T. performed the hidden-variable analysis. R.I. and T.J. produced the experimental design. R.B.M. and D.R.T. analyzed the experimental signatures. All authors contributed to the writing of the manuscript. D.R.T. coordinated the project.\\

\noindent{\bf Competing financial interests}\\
\noindent The authors declare no competing financial interests.

\end{document}